\documentclass{revtex4}

\usepackage{indentfirst}
\usepackage{amsfonts}
\usepackage{amssymb}
\usepackage[reqno,intlimits]{amsmath}
\usepackage[polish,english]{babel}

\newtheorem{theorem}{Theorem}

\begin{document}

\title{On solvable Dirac equation with polynomial potentials}
\author{Tomasz Stachowiak}
\email{toms@oa.uj.edu.pl}
\affiliation{Copernicus Center for Interdisciplinary Studies,\\
ul. Gronostajowa 3, 30-387 Krak\'ow, Poland}

\begin{abstract}
One dimensional Dirac equation is analysed with regard to the existence of
exact (or closed-form) solutions for polynomial potentials. The notion of
Liouvillian functions is used to define solvability, and it is shown that
except for the linear potentials the equation in question is not solvable.
\end{abstract}

\maketitle

\section{Introduction}

The question of solving differential equations in explicit terms has lost some
importance with the development of numerical methods, and it is also clear
that solvable cases are non-generic in the class of all physical models.
Although the linear ordinary differential equations are mostly dealt with by
series expansion or even used to define new functions, it is still instructive
if an explicit solution can be found. Obviously, exact solutions provide
deeper quantitative insight into whole classes of solutions (depending on some
parameters); and in case of perturbative approach, where the linear equation is
just the first approximation, they allow to shift the numerical approach to the
next order. Examples include Schr\"odinger like equations
\cite{Primitivo,Prim2}, and perturbation equations in cosmology \cite{TS}.

When solving the Dirac equation it usually boils down to solving a second order
linear differential equation -- be it by an appropriate change of variables, or
by a particular ansatz. Further assumptions or symmetries lead to special
solutions which can be obtained explicitly, most of which are presented in
\cite{Cooper}. Depending on the physical theory one starts with, even though
the form
of the equations stays the same, the results on solvability differ. This
happens because of various definitions of the potential -- for example, in
some supersymmetric theories the potential $W = V^2 +V'$ is taken to be a
generic polynomial, whereas here we take $V$ to be generic, which leads to
restrictions on $U$. For results on integrability of such equations see for
instance \cite{Primitivo}.

There are also many definitions of ``explicit solutions'' depending on which
special functions are considered to be simple enough. Here we will be using the
Liouvillian extensions to construct solutions. Since the equation will be of
the form
\begin{equation}
    f''(x) = r(x)f(x),
\end{equation}
where $r(x)$ is rational, the sought solutions will lie in some extension of
the field of rational functions over $\mathbb{C}$. The extension will be called
Liouvillian if it is composed of a finite number of the following steps:
\begin{enumerate}
\item Adjoining an element whose derivative is in the field.
\item Adjoining an element whose logarithmic derivative is in the field.
\item Adjoining an element algebraic in the field.
\end{enumerate}
These functions include all the elementary ones (polynomials, exponential,
logarithm), and also special cases of
transcendental functions. The reason for introducing such class is that it
naturally follows from the differential Galois theory \cite{Kaplansky} and that
for the above equation there is an algorithmic approach to checking for
solvability \cite{Kovacic}. To put it generally, when the solutions are
Liouvillian, the differential Galois group is solvable, and there exist
invariants of the equation for which one can look. This means that, at least in
theory, it is
possible to check a given class of equations and rule out the existence of any
additional exact solutions -- an advantage usually not found in other
approaches.

Thus, to put it simply, the aim of this letter is to find Liouvillian solutions
of the Dirac equation into which a polynomial potential has been incorporated.
As we will see, the conditions for such solvability are almost never met.

\section{Dirac equation with a potential}

We take the one-dimensional form of the Dirac equation
\begin{equation}
    i \partial_t\psi = (\alpha p + \beta m )\psi,
\end{equation}
where $p$ is the momentum conjugate to the coordinate $x$, i.e. $p =
-i\partial_x$, and the Dirac matrices satisfy $\alpha^2=\beta^2=1$,
$\{\alpha,\beta\}=0$.
This form follows from the standard four-dimensional form, when the bispinor
$\psi$ depends on $t$ and $x$ only. It can then be taken as a two component
vector $\psi = \left(\psi_1(t,x),\psi_2(t,x)\right)$ for simplicity. Also the
time derivative can be eliminated by the ansatz $\psi(t,x) =
\exp(-iEt)\psi(x)$, so that the equation is
\begin{equation}
    E\psi = (-i\alpha\partial_x + \beta m )\psi.
\end{equation}

The inclusion of the potential which is specified by only one scalar function
can be performed in two ways. The polynomial potential
$V=\lambda x^n+\ldots+\lambda_0$ can be the time component of a
four-vector -- like the Coulomb part of the electromagnetic field. The equation
would then become
\begin{equation}
    (V+E)\psi = (-i \alpha\partial_x + \beta m )\psi.
    \label{vcoup}
\end{equation}
Alternatively, one could introduce scalar coupling, which modifies the $m$
term to
\begin{equation}
    E\psi = \left(-i \alpha\partial_x  + \beta (V+m)\right)\psi.
    \label{scoup}
\end{equation}

In the first case, the explicit equations are
\begin{equation}
\begin{aligned}
    \psi_1' &= (m+V)\psi_1 - E\psi_2,\\
    \psi_2' &= -(m+V)\psi_2 + E\psi_1,\label{1ord}
\end{aligned}
\end{equation}
or, as second order equations,
\begin{equation}
\begin{aligned}
    \psi_1'' &= (U' + U^2 - E^2)\psi_1,\\
    \psi_2'' &= (-U' + U^2 - E^2)\psi_2,
    \label{2ord_scal}
\end{aligned}
\end{equation}
where $U=m+V$ and the Dirac matrices were taken to be
\begin{equation}
    \alpha = \left(\begin{matrix} 0 & i\\ -i & 0\end{matrix}\right),\;\;
    \beta = \left(\begin{matrix} 0 & 1\\ 1 & 0\end{matrix}\right).
\end{equation}
The vector coupling, on the other hand, gives 
\begin{equation}
\begin{aligned}
    \psi_1'' &= (iU' - U^2 + m^2)\psi_1,\\
    \psi_2'' &= (-iU' - U^2 + m^2)\psi_2,
\end{aligned}
\end{equation}
provided that
\begin{equation}
    \alpha = \left(\begin{matrix} 1 & 0\\ 0 & -1\end{matrix}\right),\;\;
\end{equation}
and $U=V+E$. It is obvious that the vector coupling can be transformed into the
scalar one with
\begin{equation}
    V \rightarrow -i V,\; E \rightarrow -i m,\; m \rightarrow i E.
    \label{corr}
\end{equation}

\section{The solutions}

Let us start with the scalar coupling.
Since the spinor components are connected through the first order equations
\eqref{1ord} it suffices to check just the first of equations
\eqref{2ord_scal}.
For polynomial $V$, and thus polynomial $U$ it is straightforward to apply the
aforementioned Kovacic algorithm. Because the only coefficient in the equation
is an even degree polynomial, only the first case to be considered in depth,
and the possible solution will be of the form $P\exp\int\omega\mathrm{d}x$,
with a monic polynomial $P$ and a rational function $\omega$.

Since the only
singular points is the infinity one needs to obtain the expansion
$[\sqrt{r}]_{\infty}$ of $r = U'+U^2-E^2$. Because only the polynomial part of
the expansion matters, it is fully and uniquely determined by comparing the
coefficients of $r$ and $[\sqrt{r}_{\infty}]^2$. Obviously, there are at most
$n+1$ terms of the expansion, and the $U^2$ term alone fixes the $n+1$ highest
terms of $r$. Thus, $U$ itself is the solution and 
\begin{equation}
    r - [\sqrt{r}_{\infty}]^2 =r-U^2=n\lambda x^{n-1} + \ldots -E^2,\label{rminus}
\end{equation}
If $n\neq1$, the algorithm gives $\omega=U$, and $\deg(P)=0$ with the additional
condition that $E^2P=0$. In other words, for $E=0$ the solution is
\begin{equation}
    \psi_1 = \exp{\int(m+V)\mathrm{d}x} = \frac{1}{\psi_2}.\label{sol1}
\end{equation}

If $n=1$, there is only the zeroth order term in \eqref{rminus} and for
$E^2/(2\lambda)\in\mathbb{N}$ we have a whole family of solutions. This is the known
case of the so-called Dirac oscillator solvable by Hermite polynomials
\cite{Cooper}.

For non-zero values of $E$ no other (non-constant) polynomial potential gives
Liouvillian solutions. Thanks to the relations \eqref{corr}, we can see that
the same holds for the vector coupling, with the appropriate interchange of $E$
and $m$. Namely, for a linear potential one has the Hermite solutions as above,
and for $m=0$ the solution is
\begin{equation}
    \psi_1 = \exp{\int i(E+V)\mathrm{d}x} = \frac{1}{\psi_2}.\label{sol2}
\end{equation}

The above results can be summarised as follows:
\begin{theorem}
    The one-dimensional Dirac equation with a polynomial potential of degree
    $n>1$ only has Liouvillian solutions when $E=0$ (scalar coupling
    \eqref{scoup}) or $m=0$ (vector coupling \eqref{vcoup}), given by formulae
    \eqref{sol1} and \eqref{sol2} respectively.

    For $n=1$ the solutions are expressible by Hermite polynomials given in
    \cite{Cooper}, and $n=0$ is equivalent to the standard Dirac equation.
\end{theorem}

Two remarks are in order. First, note that in the special cases $m=0$ ($E=0$),
the solutions given above hold for all potentials, not just polynomial ones
as can be checked by direct differentiation. Second, the practical consequences
of the outcome are that the spinor is not expressible as a polynomial function
of position, as is often the case in solvable models of quantum mechanics. This
still leaves the possibility of using new transcendental functions or series
expansion approach, but precludes the direct (explicit) computation of norms or
other general expressions involving $\psi$.

\section{Acknowledgements}
This paper was supported by grant No. N N202 2126 33 of Ministry of Science and
Higher Education of Poland.

\end{document}